\documentclass[final]{IEEEtran}
\usepackage[update,prepend]{epstopdf}

\usepackage{graphics}
\usepackage{multirow}
\usepackage{tikz}
\usepackage{bbm} 
\usepackage{pdfpages}
\usepackage{multirow}
\usepackage{subfig}
\usepackage{comment}

\usepackage{setspace}	
\usepackage{graphicx}
\usepackage{algorithm,algorithmic}
\usepackage{multicol}

\usepackage[justification=centering]{caption}
\usepackage{textcomp}
\usepackage{psfrag}
\usepackage{arydshln}
\usepackage{url}
\usepackage{soul}
\usepackage{graphicx,color}
\usepackage[nolist]{acronym}

\usepackage{mathtools,lipsum}
\usepackage{cuted}
\usepackage{amsmath}
\usepackage{graphicx}

\def\nb0{{\mathbf{0}}}
\def\nb1{{\mathbf{1}}}

\begin{document}
\graphicspath{{./Figures/}}
	\begin{acronym}

\acro{5G-NR}{5G New Radio}
\acro{3GPP}{3rd Generation Partnership Project}
\acro{ABS}{aerial base station}
\acro{AC}{address coding}
\acro{ACF}{autocorrelation function}
\acro{ACR}{autocorrelation receiver}
\acro{ADC}{analog-to-digital converter}
\acrodef{aic}[AIC]{Analog-to-Information Converter}     
\acro{AIC}[AIC]{Akaike information criterion}
\acro{aric}[ARIC]{asymmetric restricted isometry constant}
\acro{arip}[ARIP]{asymmetric restricted isometry property}

\acro{ARQ}{Automatic Repeat Request}
\acro{AUB}{asymptotic union bound}
\acrodef{awgn}[AWGN]{Additive White Gaussian Noise}     
\acro{AWGN}{additive white Gaussian noise}

\acro{APSK}[PSK]{asymmetric PSK} 

\acro{waric}[AWRICs]{asymmetric weak restricted isometry constants}
\acro{warip}[AWRIP]{asymmetric weak restricted isometry property}
\acro{BCH}{Bose, Chaudhuri, and Hocquenghem}        
\acro{BCHC}[BCHSC]{BCH based source coding}
\acro{BEP}{bit error probability}
\acro{BFC}{block fading channel}
\acro{BG}[BG]{Bernoulli-Gaussian}
\acro{BGG}{Bernoulli-Generalized Gaussian}
\acro{BPAM}{binary pulse amplitude modulation}
\acro{BPDN}{Basis Pursuit Denoising}
\acro{BPPM}{binary pulse position modulation}
\acro{BPSK}{Binary Phase Shift Keying}
\acro{BPZF}{bandpass zonal filter}
\acro{BSC}{binary symmetric channels}              
\acro{BU}[BU]{Bernoulli-uniform}
\acro{BER}{bit error rate}
\acro{BS}{base station}
\acro{BW}{BandWidth}
\acro{BLLL}{ binary log-linear learning }

\acro{CP}{Cyclic Prefix}
\acrodef{cdf}[CDF]{cumulative distribution function}   
\acro{CDF}{Cumulative Distribution Function}
\acrodef{c.d.f.}[CDF]{cumulative distribution function}
\acro{CCDF}{complementary cumulative distribution function}
\acrodef{ccdf}[CCDF]{complementary CDF}               
\acrodef{c.c.d.f.}[CCDF]{complementary cumulative distribution function}
\acro{CD}{cooperative diversity}

\acro{CDMA}{Code Division Multiple Access}
\acro{ch.f.}{characteristic function}
\acro{CIR}{channel impulse response}
\acro{cosamp}[CoSaMP]{compressive sampling matching pursuit}
\acro{CR}{cognitive radio}
\acro{cs}[CS]{compressed sensing}                   
\acrodef{cscapital}[CS]{Compressed sensing} 
\acrodef{CS}[CS]{compressed sensing}
\acro{CSI}{channel state information}
\acro{CCSDS}{consultative committee for space data systems}
\acro{CC}{convolutional coding}
\acro{Covid19}[COVID-19]{Coronavirus disease}

\acro{DAA}{detect and avoid}
\acro{DAB}{digital audio broadcasting}
\acro{DCT}{discrete cosine transform}
\acro{dft}[DFT]{discrete Fourier transform}
\acro{DR}{distortion-rate}
\acro{DS}{direct sequence}
\acro{DS-SS}{direct-sequence spread-spectrum}
\acro{DTR}{differential transmitted-reference}
\acro{DVB-H}{digital video broadcasting\,--\,handheld}
\acro{DVB-T}{digital video broadcasting\,--\,terrestrial}
\acro{DL}{DownLink}
\acro{DSSS}{Direct Sequence Spread Spectrum}
\acro{DFT-s-OFDM}{Discrete Fourier Transform-spread-Orthogonal Frequency Division Multiplexing}
\acro{DAS}{Distributed Antenna System}
\acro{DNA}{DeoxyriboNucleic Acid}

\acro{EC}{European Commission}
\acro{EED}[EED]{exact eigenvalues distribution}
\acro{EIRP}{Equivalent Isotropically Radiated Power}
\acro{ELP}{equivalent low-pass}
\acro{eMBB}{Enhanced Mobile Broadband}
\acro{EMF}{ElectroMagnetic Field}
\acro{EU}{European union}
\acro{EI}{Exposure Index}
\acro{eICIC}{enhanced Inter-Cell Interference Coordination}

\acro{FC}[FC]{fusion center}
\acro{FCC}{Federal Communications Commission}
\acro{FEC}{forward error correction}
\acro{FFT}{fast Fourier transform}
\acro{FH}{frequency-hopping}
\acro{FH-SS}{frequency-hopping spread-spectrum}
\acrodef{FS}{Frame synchronization}
\acro{FSsmall}[FS]{frame synchronization}  
\acro{FDMA}{Frequency Division Multiple Access}

\acro{GA}{Gaussian approximation}
\acro{GF}{Galois field }
\acro{GG}{Generalized-Gaussian}
\acro{GIC}[GIC]{generalized information criterion}
\acro{GLRT}{generalized likelihood ratio test}
\acro{GPS}{Global Positioning System}
\acro{GMSK}{Gaussian Minimum Shift Keying}
\acro{GSMA}{Global System for Mobile communications Association}
\acro{GS}{ground station}
\acro{GMG}{ Grid-connected MicroGeneration}

\acro{HAP}{high altitude platform}
\acro{HetNet}{Heterogeneous network}

\acro{IDR}{information distortion-rate}
\acro{IFFT}{inverse fast Fourier transform}
\acro{iht}[IHT]{iterative hard thresholding}
\acro{i.i.d.}{independent, identically distributed}
\acro{IoT}{Internet of Things}                      
\acro{IR}{impulse radio}
\acro{lric}[LRIC]{lower restricted isometry constant}
\acro{lrict}[LRICt]{lower restricted isometry constant threshold}
\acro{ISI}{intersymbol interference}
\acro{ITU}{International Telecommunication Union}
\acro{ICNIRP}{International Commission on Non-Ionizing Radiation Protection}
\acro{IEEE}{Institute of Electrical and Electronics Engineers}
\acro{ICES}{IEEE international committee on electromagnetic safety}
\acro{IEC}{International Electrotechnical Commission}
\acro{IARC}{International Agency on Research on Cancer}
\acro{IS-95}{Interim Standard 95}

\acro{KPI}{Key Performance Indicator}

\acro{LEO}{low earth orbit}
\acro{LF}{likelihood function}
\acro{LLF}{log-likelihood function}
\acro{LLR}{log-likelihood ratio}
\acro{LLRT}{log-likelihood ratio test}
\acro{LoS}{Line-of-Sight}
\acro{LRT}{likelihood ratio test}
\acro{wlric}[LWRIC]{lower weak restricted isometry constant}
\acro{wlrict}[LWRICt]{LWRIC threshold}
\acro{LPWAN}{Low Power Wide Area Network}
\acro{LoRaWAN}{Low power long Range Wide Area Network}
\acro{NLoS}{Non-Line-of-Sight}
\acro{LiFi}[Li-Fi]{light-fidelity}
 \acro{LED}{light emitting diode}
 \acro{LABS}{LoS transmission with each ABS}
 \acro{NLABS}{NLoS transmission with each ABS}

\acro{MB}{multiband}
\acro{MC}{macro cell}
\acro{MDS}{mixed distributed source}
\acro{MF}{matched filter}
\acro{m.g.f.}{moment generating function}
\acro{MI}{mutual information}
\acro{MIMO}{Multiple-Input Multiple-Output}
\acro{MISO}{multiple-input single-output}
\acrodef{maxs}[MJSO]{maximum joint support cardinality}                       
\acro{ML}[ML]{maximum likelihood}
\acro{MMSE}{minimum mean-square error}
\acro{MMV}{multiple measurement vectors}
\acrodef{MOS}{model order selection}
\acro{M-PSK}[${M}$-PSK]{$M$-ary phase shift keying}                       
\acro{M-APSK}[${M}$-PSK]{$M$-ary asymmetric PSK} 
\acro{MP}{ multi-period}
\acro{MINLP}{mixed integer non-linear programming}

\acro{M-QAM}[$M$-QAM]{$M$-ary quadrature amplitude modulation}
\acro{MRC}{maximal ratio combiner}                  
\acro{maxs}[MSO]{maximum sparsity order}                                      
\acro{M2M}{Machine-to-Machine}                                                
\acro{MUI}{multi-user interference}
\acro{mMTC}{massive Machine Type Communications}      
\acro{mm-Wave}{millimeter-wave}
\acro{MP}{mobile phone}
\acro{MPE}{maximum permissible exposure}
\acro{MAC}{media access control}
\acro{NB}{narrowband}
\acro{NBI}{narrowband interference}
\acro{NLA}{nonlinear sparse approximation}
\acro{NLOS}{Non-Line of Sight}
\acro{NTIA}{National Telecommunications and Information Administration}
\acro{NTP}{National Toxicology Program}
\acro{NHS}{National Health Service}

\acro{LOS}{Line of Sight}

\acro{OC}{optimum combining}                             
\acro{OC}{optimum combining}
\acro{ODE}{operational distortion-energy}
\acro{ODR}{operational distortion-rate}
\acro{OFDM}{Orthogonal Frequency-Division Multiplexing}
\acro{omp}[OMP]{orthogonal matching pursuit}
\acro{OSMP}[OSMP]{orthogonal subspace matching pursuit}
\acro{OQAM}{offset quadrature amplitude modulation}
\acro{OQPSK}{offset QPSK}
\acro{OFDMA}{Orthogonal Frequency-division Multiple Access}
\acro{OPEX}{Operating Expenditures}
\acro{OQPSK/PM}{OQPSK with phase modulation}

\acro{PAM}{pulse amplitude modulation}
\acro{PAR}{peak-to-average ratio}
\acrodef{pdf}[PDF]{probability density function}                      
\acro{PDF}{probability density function}
\acrodef{p.d.f.}[PDF]{probability distribution function}
\acro{PDP}{power dispersion profile}
\acro{PMF}{probability mass function}                             
\acrodef{p.m.f.}[PMF]{probability mass function}
\acro{PN}{pseudo-noise}
\acro{PPM}{pulse position modulation}
\acro{PRake}{Partial Rake}
\acro{PSD}{power spectral density}
\acro{PSEP}{pairwise synchronization error probability}
\acro{PSK}{phase shift keying}
\acro{PD}{power density}
\acro{8-PSK}[$8$-PSK]{$8$-phase shift keying}
\acro{PPP}{Poisson point process}
\acro{PCP}{Poisson cluster process}
 
\acro{FSK}{Frequency Shift Keying}

\acro{QAM}{Quadrature Amplitude Modulation}
\acro{QPSK}{Quadrature Phase Shift Keying}
\acro{OQPSK/PM}{OQPSK with phase modulator }

\acro{RD}[RD]{raw data}
\acro{RDL}{"random data limit"}
\acro{ric}[RIC]{restricted isometry constant}
\acro{rict}[RICt]{restricted isometry constant threshold}
\acro{rip}[RIP]{restricted isometry property}
\acro{ROC}{receiver operating characteristic}
\acro{rq}[RQ]{Raleigh quotient}
\acro{RS}[RS]{Reed-Solomon}
\acro{RSC}[RSSC]{RS based source coding}
\acro{r.v.}{random variable}                               
\acro{R.V.}{random vector}
\acro{RMS}{root mean square}
\acro{RFR}{radiofrequency radiation}
\acro{RIS}{Reconfigurable Intelligent Surface}
\acro{RNA}{RiboNucleic Acid}
\acro{RRM}{Radio Resource Management}
\acro{RUE}{reference user equipments}
\acro{RAT}{radio access technology}
\acro{RB}{resource block}

\acro{SA}[SA-Music]{subspace-augmented MUSIC with OSMP}
\acro{SC}{small cell}
\acro{SCBSES}[SCBSES]{Source Compression Based Syndrome Encoding Scheme}
\acro{SCM}{sample covariance matrix}
\acro{SEP}{symbol error probability}
\acro{SG}[SG]{sparse-land Gaussian model}
\acro{SIMO}{single-input multiple-output}
\acro{SINR}{signal-to-interference plus noise ratio}
\acro{SIR}{signal-to-interference ratio}
\acro{SISO}{Single-Input Single-Output}
\acro{SMV}{single measurement vector}
\acro{SNR}[\textrm{SNR}]{signal-to-noise ratio} 
\acro{sp}[SP]{subspace pursuit}
\acro{SS}{spread spectrum}
\acro{SW}{sync word}
\acro{SAR}{specific absorption rate}
\acro{SSB}{synchronization signal block}
\acro{SR}{shrink and realign}

\acro{tUAV}{tethered Unmanned Aerial Vehicle}
\acro{TBS}{terrestrial base station}

\acro{uUAV}{untethered Unmanned Aerial Vehicle}
\acro{PDF}{probability density functions}

\acro{PL}{path-loss}

\acro{TH}{time-hopping}
\acro{ToA}{time-of-arrival}
\acro{TR}{transmitted-reference}
\acro{TW}{Tracy-Widom}
\acro{TWDT}{TW Distribution Tail}
\acro{TCM}{trellis coded modulation}
\acro{TDD}{Time-Division Duplexing}
\acro{TDMA}{Time Division Multiple Access}
\acro{Tx}{average transmit}

\acro{UAV}{Unmanned Aerial Vehicle}
\acro{uric}[URIC]{upper restricted isometry constant}
\acro{urict}[URICt]{upper restricted isometry constant threshold}
\acro{UWB}{ultrawide band}
\acro{UWBcap}[UWB]{Ultrawide band}   
\acro{URLLC}{Ultra Reliable Low Latency Communications}
         
\acro{wuric}[UWRIC]{upper weak restricted isometry constant}
\acro{wurict}[UWRICt]{UWRIC threshold}                
\acro{UE}{User Equipment}
\acro{UL}{UpLink}

\acro{WiM}[WiM]{weigh-in-motion}
\acro{WLAN}{wireless local area network}
\acro{wm}[WM]{Wishart matrix}                               
\acroplural{wm}[WM]{Wishart matrices}
\acro{WMAN}{wireless metropolitan area network}
\acro{WPAN}{wireless personal area network}
\acro{wric}[WRIC]{weak restricted isometry constant}
\acro{wrict}[WRICt]{weak restricted isometry constant thresholds}
\acro{wrip}[WRIP]{weak restricted isometry property}
\acro{WSN}{wireless sensor network}                        
\acro{WSS}{Wide-Sense Stationary}
\acro{WHO}{World Health Organization}
\acro{Wi-Fi}{Wireless Fidelity}

\acro{sss}[SpaSoSEnc]{sparse source syndrome encoding}

\acro{VLC}{Visible Light Communication}
\acro{VPN}{Virtual Private Network} 
\acro{RF}{Radio Frequency}
\acro{FSO}{Free Space Optics}
\acro{IoST}{Internet of Space Things}

\acro{GSM}{Global System for Mobile Communications}
\acro{2G}{Second-generation cellular network}
\acro{3G}{Third-generation cellular network}
\acro{4G}{Fourth-generation cellular network}
\acro{5G}{Fifth-generation cellular network}	
\acro{gNB}{next-generation Node-B Base Station}
\acro{NR}{New Radio}
\acro{UMTS}{Universal Mobile Telecommunications Service}
\acro{LTE}{Long Term Evolution}

\acro{QoS}{Quality of Service}
\end{acronym}
	
\newcommand{\SAR} {\mathrm{SAR}}
\newcommand{\WBSAR} {\mathrm{SAR}_{\mathsf{WB}}}
\newcommand{\gSAR} {\mathrm{SAR}_{10\si{\gram}}}
\newcommand{\Sab} {S_{\mathsf{ab}}}
\newcommand{\Eavg} {E_{\mathsf{avg}}}
\newcommand{\ft}{f_{\textsf{th}}}
\newcommand{\alphatf}{\alpha_{24}}

\title{
HAPS in the Non-Terrestrial Network Nexus:
Prospective Architectures and Performance Insights
}
\author{
Zhengying Lou, Baha Eddine Youcef Belmekki,~\IEEEmembership{Member,~IEEE}, and Mohamed-Slim Alouini,~\IEEEmembership{Fellow,~IEEE}
\thanks{The authors are with King Abdullah University of Science and Technology (KAUST), CEMSE division, Thuwal 23955-6900, Saudi Arabia. E-mail: \{zhengying.lou,bahaeddine.belmekki,slim.alouini\}@kaust.edu.sa.}
\vspace{-4mm}
}
\maketitle
\thispagestyle{empty}
 \pagestyle{empty}

\begin{abstract}
 High altitude platform stations (HAPS) have recently emerged as a new key stratospheric player in non-terrestrial networks (NTN) alongside satellites and low-altitude platforms. In this paper, we present the main communication links between HAPS and other NTN platforms, their advantages, and their challenges. Then, prospective network architectures in which HAPS plays an indispensable role in the future NTNs are presented such as ad-hoc, cell-free, and integrated access and backhaul. To showcase the importance of HAPS in the NTN, we provide comprehensive performance insights when using HAPS in the prospective architectures with the most suitable communication link. 
The insights show the HAPS' ability to interconnect the NTN nexus as well as their versatility by incorporating different metrics into the analysis
such as routing latency, energy efficiency, coverage probability, and channel capacity.
Depending on the architecture, HAPS will play different roles in NTN, such as a UAV network center, satellite relay, and ground network extension.
Finally, the performance gain provided by HAPS usage in NTN is further highlighted by comparing the results when no HAPS are used.

\end{abstract}


\vspace{-0.2cm}

\section{Introduction}
Unprecedented traffic and processing needs as well as the exponential growth in smart devices are driving the growth of non-terrestrial networks (NTNs). 
According to the third generation partnership project (3GPP), NTN involves all flying objects, including unmanned aerial vehicles (UAVs), high altitude platform stations (HAPS), and satellites \cite{huang2023system,cao2018airborne}. \textcolor{black}{\cite{huang2023system} offers a comprehensive analysis of system-level metrics for NTN, utilizing stochastic geometry as a system-level analysis tool. This research also includes a detailed discussion of communication issues and application scenarios for NTN, demonstrating the importance and potential performance gains of NTN.}
Specifically, NTN platforms can provide ubiquitous wireless coverage in areas where fiber is too expensive or difficult to deploy such as remote rural areas, oceans, and mountains. 
Compared with fiber optics, wireless links bring significant cost reductions while achieving approximate data transmission rates. In addition, NTN platforms such as HAPS and UAVs can be quickly deployed owing to their high flexibility and mobility  \cite{cao2018airborne}. Therefore, NTN are widely applied to ease disconnection problems in post-disaster networks and improve rescue efficiency \cite{belmekki2022unleashing}. Moreover, NTN can serve as an extension of the existing ground network, enhancing the capacity of the existing ground network and reducing communication latency \cite{wang2022stochastic}. Therefore, they are expected to serve future aerial users and platforms such as electric vertical take-off and landing (eVTOL) aircraft in urban air mobility \cite{zaid2021evtol}.


\par
HAPS have drawn a lot of attention during the last years; and although the concept of HAPS was proposed more than twenty years ago, it did not play an important role in NTN due to the challenges related to stratospheric conditions, materials, and energy. \textcolor{black}{The platforms in the stratospheric suffer from harsh flying conditions, such as low temperatures of -65$^{\circ}$C, wind speeds exceeding 40~km/h, and prolonged solar radiation \cite{belmekki2022unleashing}.} However, with the development of lightweight composite materials, as well as the improvement of solar panel efficiency and antenna technology, there have been numerous technological research and commercial application projects. \textcolor{black}{about HAPS such as Google Loon. Although the project was dropped, it paved the way for these new research for stratospheric connectivity. In 2022, the first world-ever 5G trial using HAPS was performed in the red sea region, providing a coverage of 450 km$^2$ with data rates approximating 90 Mbps.}
\par
HAPS add several benefits to NTN due to their high performance and critical positioning. HAPS outperform satellites in terms of link capacity, launch and maintenance costs, and transmission latency \cite{benyahia2022haps}. Compared to UAVs, HAPS can keep quasi-static relative to the ground for longer periods, carry more communication units, and offer a broader range of services. 
{\color{black} HAPS can be classified into two categories based on the physical principle that provides their lifting force. Aerostatic, that is, lighter than air platforms, rely on buoyancy to remain airborne and have a high payload (tens to hundreds of kilograms).  
On the other hand, aerodynamic, that is, heavier than air platforms, generate dynamic forces through movement through the air, granting them superior maneuverability.
 }
In addition, massive HAPS constellations--in contrast to satellites and UAVs--have not yet been deployed, and there are currently insufficient support facilities and regulatory policies pertaining to HAPS. 

The ultra-densification of the cellular ground, the near-ground, and the constellations of tens thousands satellites that are expected to be deployed necessitates the exploration for another dimension of connectivity. Therefore, exploiting the stratosphere allows to leverage an unused layer with a large gamut of applications. In that regard, we investigate the impact of HAPS to interconnect the NTN nexus and demonstrate how HAPS can effectively improve and complement connectivity in NTN.

\section{Transmission Links and Frequency Bands} \label{section2}

\subsection{HAPS-HAPS Link}
Most of the HAPS-HAPS links in the literature are utilized for long-distance backhaul and require high transmission rates, such as in emergency response and remote monitoring. Free space optics (FSO), which can achieve tens of Gbps data rate and very low latency, is a commonly used frequency band in HAPS-HAPS links. However, FSO signal requires precise alignment between the transmitter and receiver, which can be challenging to maintain over long distances \cite{trichili2020roadmap}. In addition, the atmospheric turbulence caused by the temporal and spatial inhomogeneity of stratospheric temperature and pressure also interferes with FSO communication \cite{benyahia2022haps}. As solutions, advanced beam steering and tracking techniques can be used to maintain precise alignment, also FSO can be combined with other communication technologies, like millimeter wave (mmWave), to provide a backup communication link. {\color{black} Nonetheless, FSO communication equipment is typically bulky, often exceeding $5$ kg in weight. As a result, this equipment is suitable for deployment on aerostatic HAPS designed to accommodate substantial payloads.}
\par 
In addition to FSO communication, radio communications with high-frequency bands, such as mmWave and terahertz (THz) \cite{lou2023coverage}, are also optional communication modes for HAPS-HAPS links, as they can deliver high data rates and low latency due to the wide available bandwidth. However, mmWave and THz signals suffer from substantial path loss, which is proportional to the square of frequency and the square of the distance, as well as atmospheric absorption, which is more severe at higher frequencies. To address these challenges, directional antennas with high gain and beamforming techniques can be utilized to reduce path loss and enhance the link budget. {\color{black} Furthermore, high-frequency antennas, due to their reduced wavelengths, can be designed to be compact, effectively addressing concerns about their size.} In addition, adaptive power control and dynamic spectrum access can be used to optimize the utilization of available spectrum and reduce interference from other systems operating in the same frequency band.

\begin{figure*}[th]
	\centering
	\includegraphics[width=0.98\linewidth]{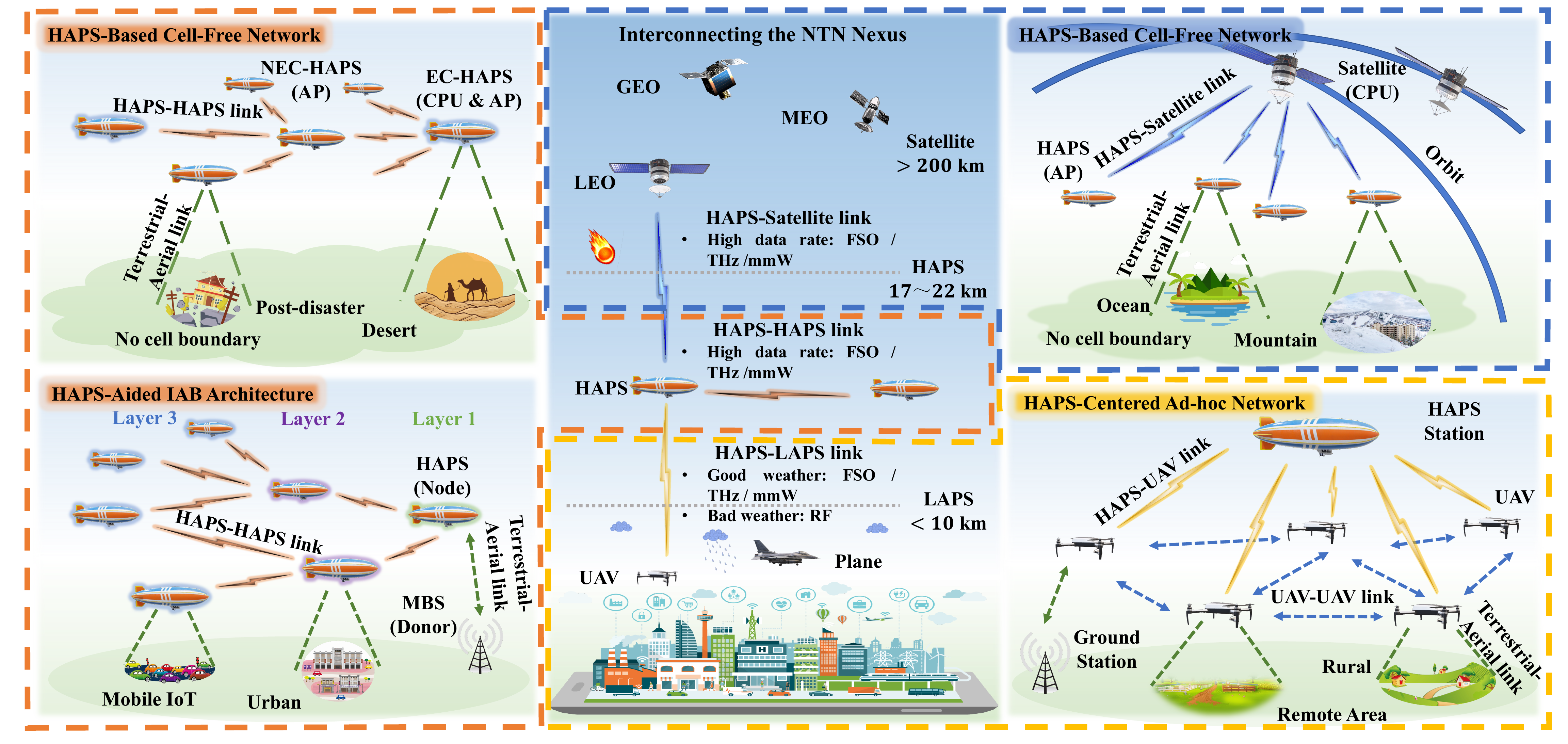}
	\caption{System framework diagram.}
	\label{fig:big}
	\vspace{-4mm}
\end{figure*}

\subsection{HAPS-Satellite Link}
\label{sec2-2}

HAPS can typically establish backhaul links with LEO satellites to provide connectivity for remote areas or be deployed as a relay between the satellite and the mobile platform to enhance the communication performance \cite{benyahia2022haps}. Similar to HAPS-HAPS links, FSO and high-frequency radio communications are still the preferred communication modes for HAPS-satellite backhaul links. Regarding FSO, the factors influencing the quality of HAPS-HAPS links, such as pointing error, atmospheric attenuation, and turbulence, still affect HAPS-satellite links. And for radio communication, the path loss is greater since the length of a HAPS-satellite link is significantly longer than an inter-HAPS link.
\par
Unlink HAPS that are quasi-stationary, the high-speed movement of LEO satellites poses a challenge for the HAPS-satellite links, whether using FSO or high-frequency radio communications. The Doppler effect caused by the high relative velocity can lead to significant frequency shift and signal attenuation, which can degrade link performance \cite{trichili2020roadmap}. Adaptive optics that can adjust the pointing and tracking of the laser beam to compensate for the beam wander caused by the satellite movement is a solution for FSO links. For high-frequency radio communication, diversity techniques such as space-time coding and beamforming can alleviate the signal degradation caused by satellite motion.

\subsection{HAPS-UAV Link}
\label{sec2-3}

HAPS typically establish access links with UAVs, such as delivering real-time video surveillance during military operations or real-time data transmission for irrigation control in agricultural monitoring and management. In a stable and clear weather, HAPS-UAV links can use FSO or high-frequency radio communication, both of which offer low latency and interference resistance. Since HAPS-UAV links traverse the troposphere in addition to the stratosphere, for high-frequency radio signals, the energy loss induced by molecular vibration is an important additional issue impacting the communication quality of HAPS-UAV links, particularly in tropical rainforests. As for FSO signals, absorption and scattering of dust, fog, and clouds in the troposphere will cause greater atmospheric attenuation than in the stratosphere. Using hybrid communication systems, which combine these technologies with radio frequency (RF), can provide a reliable and high-rate communication solution.
\par
The RF transmissions below $6$~GHz can be utilized when transmission reliability requirements are high or when the transmission environment is harsh \cite{abbasi2022uxnb}. RF signals do not undergo significant atmospheric attenuation and turbulence in the troposphere, but the achievable data rate is relatively low, and the interference is significant. The application of frequency hopping, directional antennas, and spread spectrum techniques can reduce interference and enhance signal quality. \textcolor{black}{Advanced modulation techniques such as quadrature amplitude modulation (QAM), enable the transmission of more data per unit of bandwidth, therefore, enhancing the transmission rate.}

\section{Prospective Architectures}  \label{section3}

\subsection{HAPS-Centered Ad-hoc Network}
Traditional flying ad-hoc networks (FANETs) consist of several UAVs connected in an ad-hoc fashion. One of the advantages of FANETs is their ability to quickly deploy and operate in remote or disaster areas. FANETs can also provide flexible and scalable communication links that can adapt to changing environmental conditions and network topologies. However, the equal status of UAVs in FANETs, which means that all UAVs have the same capabilities and responsibilities without any hierarchical differentiation, poses significant challenges to robust control and routing stability\cite{bekmezci2013flying}. In FANETs, UAVs must frequently interact with nearby UAVs, also known as "neighbors," which may appear on their predicted trajectories to avoid potential collisions. The trajectory adjustment of any UAV might change the scheduling of the entire network. In addition, routing also faces the same aforementioned challenge. 
\par
{\color{black}  To address the limitations of FANETs, HAPS have been introduced as a solution, with aerostatic HAPS capable of carrying exceedingly powerful processors in particular.}
 In a HAPS-centered ad-hoc network, HAPS is required to establish communication links with all the UAVs, as shown in the bottom right of Fig.~\ref{fig:big}. With these connections, UAVs can share global information in HAPS-centered ad-hoc networks, as opposed to simply acquiring local information in FANETs \cite{abbasi2022uxnb}. Consequently, a HAPS-centered ad-hoc network can mitigate the risk of distributed control and the resource waste of distributed routing. Furthermore, these links are generally short in distance and within line-of-sight (LoS), which means that the communication delay is small and the energy consumption is low. As discussed in Sec. \ref{sec2-3}, the communication between HAPS and UAVs that occurs through the troposphere presents reliability and security issues. To overcome these issues, a multi-path transmission system can improve communication dependability, while an encryption technique can ensure communication security.

\subsection{HAPS-Based Cell-Free Network}
A cell-free network consists of central processing units (CPUs) and access points (APs) as shown in the right top of Fig.~\ref{fig:big} \cite{elhoushy2021cell}. In contrast to a cellular network, where each communication AP serves a specific area called a cell, a cell-free network utilizes multiple APs without explicit cell boundaries to improve coverage, capacity, and energy efficiency. The link conditions of HAPS and satellite communication, described in Sec. \ref{sec2-2}, pose certain challenges due to long-distance transmission and the fast movement of LEO satellites. By optimizing the path and protocol for signal transmission between HAPS and satellites, network latency can be reduced. In addition, implementing self-organizing network technology permits nodes to reorganize and coordinate dynamically, adjusting to changes in network topology and signal, hence, boosting network performance and dependability.
\par
Non-edge computing (NEC)-HAPS and edge computing (EC)-HAPS combined cell-free network is a relatively new architecture \cite{ke2021edge}. As shown in the top left of Fig.~\ref{fig:big}, the NEC-HAPS and EC-HAPS correspond to APs and CPUs in the cell-free network, respectively. 
{\color{black} NEC-HAPS are designed as aerial APs to receive and transmit signals, requiring only a simple communication antenna and radio frequency chains; thus, aerodynamic HAPS with lower costs and superior mobility control are optional. Onboard EC-HAPS edge servers can be used as processing units to perform a variety of computing tasks, including complex signal detection, massive data processing, and information fusion, necessitating the use of aerostatic HAPS with large payload capacities and high energy generation capabilities. }
Each NEC-HAPS will connect to multiple EC-HAPS when a computational task is required. Data can be quickly processed and transmitted back to the user through the collaboration of multiple EC-HAPS. In addition, effective cooperation between HAPS can lower the expense of covering the same area. \textcolor{black}{In this network, edge devices may pose security issues, such as malicious assaults and hacker intrusions. Therefore, it is necessary to implement security methods such as encrypted communication and identity authentication to ensure network security.}

\subsection{HAPS-Aided Integrated Access and
Backhaul Architecture}
Prior to the maturation of 5G technologies, there was little commercial interest in using a portion of the spectrum specifically for backhaul transmission. With the large bandwidth of 5G and beyond, spectrum resources can be separated into backhaul and access. 
This spectrum division technology enables the implementation of a hierarchical architecture known as integrated access and backhaul (IAB)
\cite{zhang2021survey}. 
\par
A HAPS-aided IAB architecture and its devices are shown in the bottle middle of Fig.~\ref{fig:big}, and devices in IAB architecture are classified into two types: IAB donors and IAB nodes \cite{zhang2021survey}. {\color{black}  IAB donors typically consist of terrestrial network devices such as fiber-wired macro-cell base stations (MBSs), that are wired and connected to the core network. Moreover,  IAB nodes consist of devices in NTNs, serving as extensions of the ground network and accessing the network via wireless links.  In exceptional cases, such as post-disaster situations or remote areas without ground communication equipment, NTNs may act as donors and be directly connected to the core network.} IAB donors are at the lowest level of this hierarchical architecture, while the upper layers are composed of IAB nodes that include HAPS. 
More resource units (time/frequency slots) are assigned to IAB nodes as their layer position decreases.
Although IAB nodes at each layer are intended to provide the same coverage, the throughput of nodes at lower layers is larger due to the different number of resource blocks occupied. 
\par
{\color{black} Since the IAB network only supports the mobility of the last hop, the other wireless backhaul links are static, thus, the locations of most nodes need to be relatively fixed \cite{zhang2021survey}. In this context, aerostatic HAPS emerge as an optimal choice due to their ability to remain airborne for extended periods using buoyancy or solar energy, ensuring exceptional stability and reliability. For temporary hotspots such as concerts and sports events, the last hop can leverage the greater flexibility of aerodynamic HAPS.} \textcolor{black}{However, massive deployment of HAPS networks currently lacks sufficient regulation, which may result in potential interference with other wireless communication systems, as well as safety and privacy concerns. To address these issues, it is important to establish clear regulations and standards for HAPS network deployment, including frequency allocation, power levels, and safety measures.}


\section{Interconnecting the 
NTN Nexus: Metrics, Simulations, and Insight} \label{section4}

  
In order to showcase the effectiveness of HAPS, we provide comprehensive numerical results for all the architecture and for different communication links. Therefore, we show how HAPS can interconnect the NTN nexus according to several performance metrics.
Depending on the architecture, HAPS will play different roles in NTN, such as a UAV network center, satellite relay, and ground network extension. 
We choose suitable transmissions for different interconnecting links, and we model the position of the user and aerial platform based on common stochastic geometric models. 

\begin{table}[h]
\centering
\caption{Summary of the simulation parameters.}
\label{table}
\resizebox{\linewidth}{!}{
 \renewcommand{\arraystretch}{1.1}
\scalebox{0.75}{\begin{tabular}{|c|c|}
\hline 
Parameter                                  & Value                              \\ \hline  \hline 
Altitude of user~/~UAV~/~HAPS~/~satellite              & 0~/~0.05~/~20~/~550~[km]                  \\ \hline
Antenna gain of user~/~UAV~/HAPS~/~satellite          & 3~/~10~/~30~/~50~[dBi]               \\ \hline
Transmit power of user~/~UAV~/~HAPS~/~satellite        & 20~/~30~/~36~/~45~[dBm]               \\ \hline
Carrier frequency of RF~/~mmWave link           & 2~/~28~[GHz]                       \\ \hline
Bandwidth of RF~/~mmWave link                   & 40~/~100~[MHz]                     \\ \hline
Shape parameter of  shadowed-Rician fading & (1.29, 0.158, 19.4)                \\ \hline
Shape parameter of Nakagami-$m$ fading       & 2                          \\ \hline
Number of antenna elements of cosine antenna pattern       &  32                         \\ \hline
Noise power                                & -174~dBm/Hz                         \\ \hline Packet size                                & 5 Mbits                       \\ \hline 
\end{tabular}}}
\end{table}

\subsection{Routing Latency in Ad-hoc Networks}\label{simulation1}


\subsubsection{Metric} 
We first investigate the impact of HAPS availability and routing strategies on routing latency in ad-hoc networks are analyzed. 
\textcolor{black} {Routing latency is the sum of propagation latency and transmission latency. Propagation latency represents the time required for a signal to traverse through the medium, while transmission latency is defined as the time required for a packet to be transmitted at the maximum achievable rate.}
We note that the transmission latency can be calculated as \textcolor{black}{packet} size divided by channel capacity, and the channel capacity can be obtained by the average capacity given by Shannon's theorem over a fading channel \cite{wang2022ultra}. 
As mentioned, HAPS can mitigate the risk of distributed control and avoid blockage. Routing latency is a pertinent performance measure for HAPS-centered ad-hoc networks since it reflects the performance degradation due to the increase in communication distance. It also reflects the performance gain when direct links are possible that is, if the UAV detects that the next hop is obstructed, it will send the information to the HAPS. Then, the HAPS will forward the information directly to the receiver.

\par

\subsubsection{System Model}
We consider $1000$ UAVs distributed on a disc with a radius of $20$~km forming a homogeneous binomial point process (BPP). 
{\color{black} When HAPS is available, an aerostatic HAPS with powerful processors locates in the center of the disc to assist communication between UAVs.}
{\color{black} An RF transmission is used for both UAV-UAV and HAPS-UAV links, and its parameters are provided in Table~\ref{table}. As for small-scale fading, UAV-UAV and HAPS-UAV channels follow Nakagami-$m$ fading and shadowed Rician fading, respectively. Furthermore, we assume the channel model given in \cite{wang2022stochastic} also applies to the HAPS-UAV channel. Considering that ad-hoc networks are usually used to provide communication between relief workers and residents of disaster areas when cellular infrastructures are damaged, the UAV-UAV channel located in urban environments obeys the model proposed in \cite{alzenad2019coverage}, and the only difference is the blockage probability.}

The UAVs in the network can avoid being blocked by solid obstacles such as buildings by using scheduling algorithms, however, they cannot avoid soft obstacles such as trees. The link under vegetarian blockage is called the obstructed line-of-sight (OLoS) link, which suffers an extra $20$dB path loss compared to the LoS link. Assuming that the distribution of trees in the region forms a homogeneous Poisson point process (PPP), the probability of establishing LoS links among UAVs follows an exponential distribution with a parameter of 0.08. The distance between the UAVs is the independent variable in this distribution, meaning that the probability of establishing a LoS link decreases exponentially with increasing distance.
\par

\subsubsection{Communication Mechanism}
In what follows, we detail three communication mechanisms and compare their performance.
\begin{itemize}
    \item \textbf{Long hop and HAPS not available}: UAV chooses the nearest one to the receiver (destination) as the next hop. Furthermore, the next hop should be within a communication range, which is set to $10$~km in this case. Note that when the receiver is within the communication range, the routing process concludes by transmitting the \textcolor{black}{packet} to the receiver.
    \item \textbf{Short hop and HAPS not available}: According to the short hop strategy proposed in \cite{wang2022stochastic}, the UAV will choose the nearest neighbor within the range of directional angle as the next hop.
    \item \textbf{HAPS available}: Considering the above two strategies, if the UAV detects that the next hop is obstructed, it will send the information to the HAPS. Then, the HAPS will forward the information directly to the receiver.
\end{itemize}

\begin{figure}[h]
	\centering
	\includegraphics[width=0.8\linewidth]{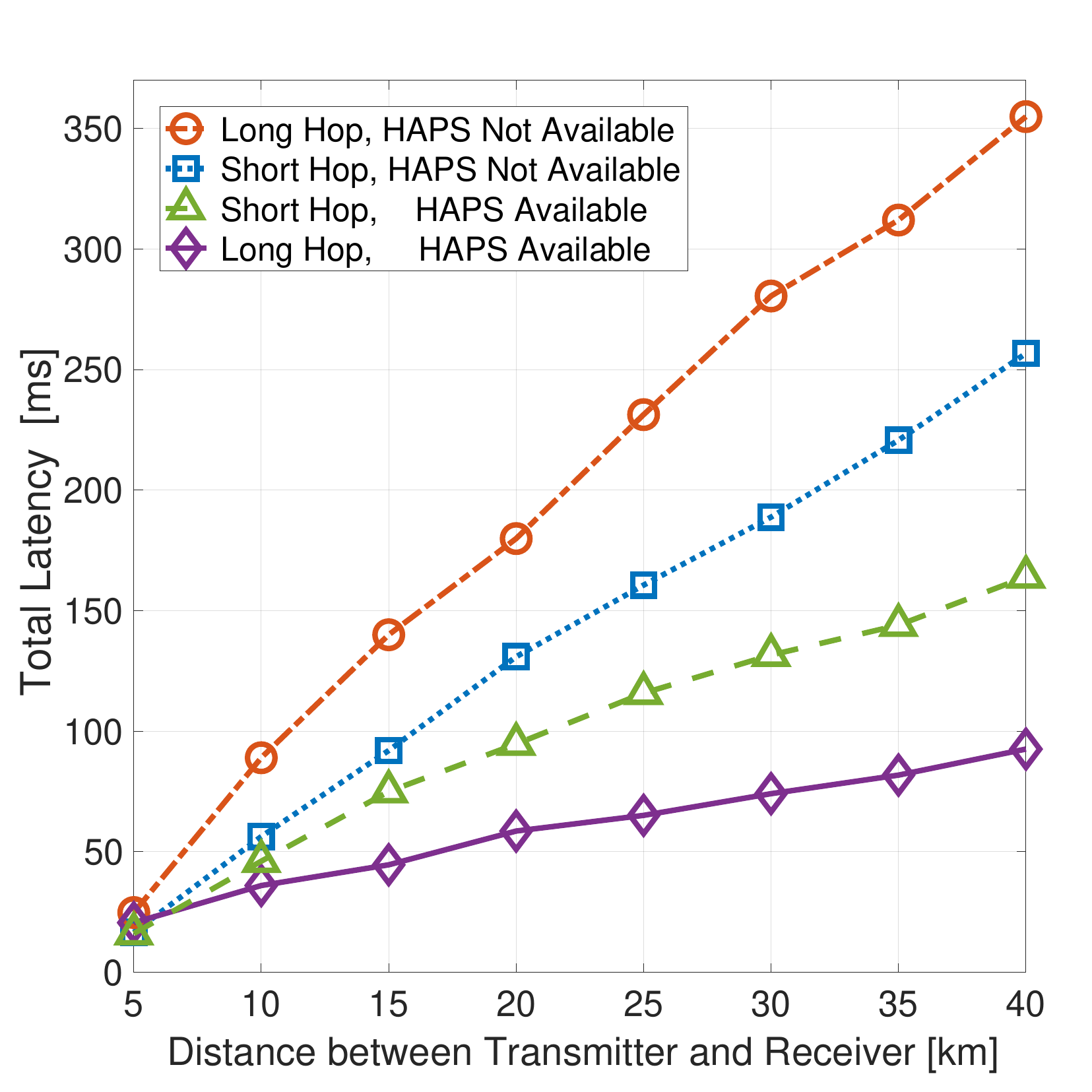}
	\caption{The impact of HAPS availability and routing strategies on routing latency.}
	\label{fig:Ad-hoc}
\end{figure}

\subsubsection{Insights}
 As shown in Fig.~\ref{fig:Ad-hoc}, the total latency increases when the distance between the transmitter and receiver increases.  The larger the distance, the greater the gain of having a HAPS located in the center of the disk for latency reduction.
 When the packet size is $5$~Mbits, the transmission latency is greater than the propagation latency. Note that a larger packet takes more time to transmit, while the signal propagation latency is mainly dependent on the total propagation distance.
When HAPS is available, the transmitter or relay UAV could send the packet to the receiver via a HAPS used as a relay, while the blockage probability increases exponentially with the transmission distance. Compared to waiting for several short hops until the message is blocked, having HAPS assistance during long hop transmissions can better meet low-latency requirements.
When HAPS is not available, the total latency of the long hop strategy is significantly longer than the short-hop strategy. This is because the blocking probability and signal attenuation increase with the increase of transmission distance, so the transmission capacity of long hop is limited. 
The findings in this study suggest that using HAPS as a relay in UAV ad-hoc network can be particularly useful in emergency response situations, such as post-disaster search and rescue missions, where low-latency communication is critical for coordinating and carrying out rescue efforts efficiently. 



\subsection{Energy Analysis in HAPS-Centered Cell-Free Networks} \label{simulation2}

\subsubsection{Metric}
Energy efficiency is defined as the number of bits transmitted per unit of energy consumed. It can also be defined as the ratio of channel capacity and transmission power.

\subsubsection{System Model}
The UAVs and HAPS act as APs and CPUs respectively, and they provide service for ground users. The UAVs and HAPS, following the BPP, are distributed on discs with a radius of $50$~km at different altitudes. Users form a homogeneous PPP with a density of $1$~km$^{-2}$, while one-tenth of them are active users who are simultaneously involved in communication.  {\color{black} 
 Both UAV-user and UAV-HAPS links use mmWave transmission, with parameters detailed in Table~\ref{table}. Additionally, a cosine antenna pattern and $10$ orthogonal sub-bands have been implemented to reduce interference and boost energy efficiency.
 The channel model of the HAPS-UAV links is similar to the ones used in \cite{wang2022stochastic}. The channel model of the UAV-user link is the same as the channel model proposed in \cite{alzenad2019coverage}, and the blockage probability is a function of the elevation angle formed between the UAV to the user.}
\par

\subsubsection{Communication Mechanism}
In cellular networks, a user is associated with the closest UAV, whether the UAV is blocked or not. In the cell-free network, the user is associated with the UAV that provides the strongest received power. Since the HAPS-UAV link is always in LoS, the UAV will be associated with the nearest HAPS in both architectures for data transmission. The overall energy efficiency of the two links is equal to the product of their individual energy efficiencies divided by the sum of their individual energy efficiencies.
\par

\begin{figure}[h]
	\centering
	\includegraphics[width=0.8\linewidth]{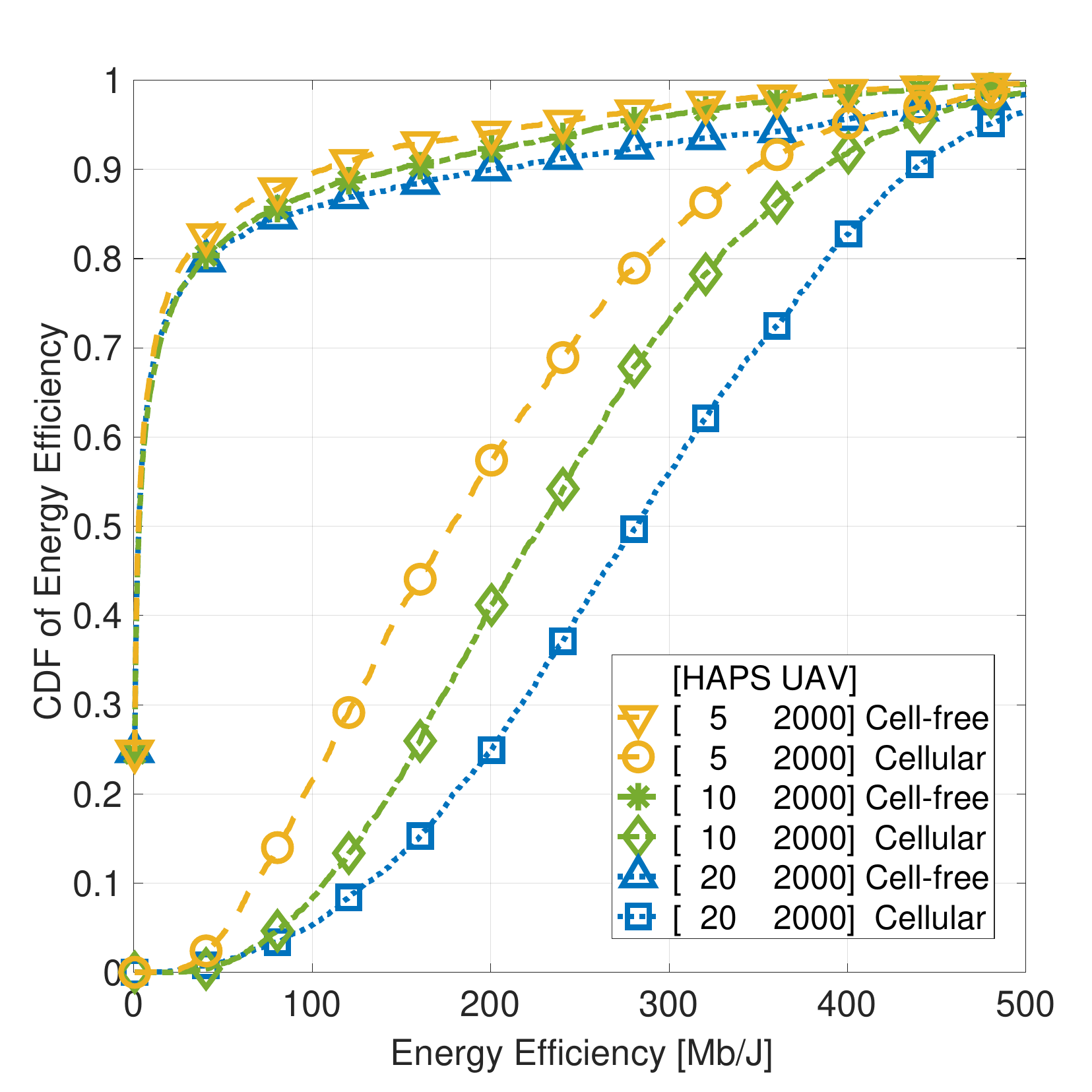}
	\caption{Effect of cell-free architecture on energy efficiency under different number of HAPS.}
	\label{fig:energy}
\end{figure}


\subsubsection{Insights}
{\color{black} 
Fig.~\ref{fig:energy} shows the cumulative distribution function (CDF) of energy efficiency. In the cellular network, more than $80\%$ of transmissions cannot achieve an energy efficiency of $40$~Mb$/$J, while nearly every communication in the cell-free network can exceed $40$~Mb/J and the average energy efficiency can reach $250$~Mb$/$J. In addition to the architecture of the network, the number of HAPS also has a slight impact on energy efficiency. Increasing the number of HAPS can reduce the average communication distance between a UAV and its associated HAPS, hence enhancing SINR and improving energy efficiency. The results indicate that the UAV cell-free network architecture with an adequate number of HAPS could be a promising strategy for enhancing energy efficiency.}

\subsection{Coverage Analysis in HAPS-Relayed Cell-Free Networks} \label{simulation3}


\subsubsection{Communication Mechanism}
The cell-free network considered is composed of satellites and HAPS. When HAPS is available, users rely on HAPS for access, and HAPS uploads data to satellites. When HAPS is unavailable, the satellite acts as an AP and CPU at the same time.

\subsubsection{Metric}
{\color{black} Unlike the UAV-user link, which is only tens to hundreds of meters away, the satellite-user link may not be as stable due to the satellite's rapid movements and considerable communication distance. Therefore, coverage probability, which is defined as the probability that \ac{SINR} of the communication link is greater than a certain threshold, is a more relevant metric than energy efficiency in HAPS-relayed cell-free networks. Note that when HAPS acts as a relay between the user and the satellite, successful coverage necessitates that the SINR of both the HAPS-user and satellite-HAPS links must exceed the coverage requirement.}
\par

\subsubsection{System Model}
The distribution of user and HAPS follow the same model as in subsection~\ref{simulation2},  and satellites form a homogeneous BPP on the whole sphere \cite{wang2022ultra}. {\color{black} The HAPS/satellite-user and the satellite-HAPS links employ mmWave transmission with the parameters specified in Table~\ref{table}, and also utilize orthogonal sub-bands and narrow-beam transmission. In addition, the channel model of the HAPS/satellite-user link is similar to the one used in  \cite{wang2022stochastic}, whereas the HAPS-satellite link follows the free space channel model without small-scale fading.}

\par

\begin{figure}[ht]
	\centering
	\includegraphics[width=0.8\linewidth]{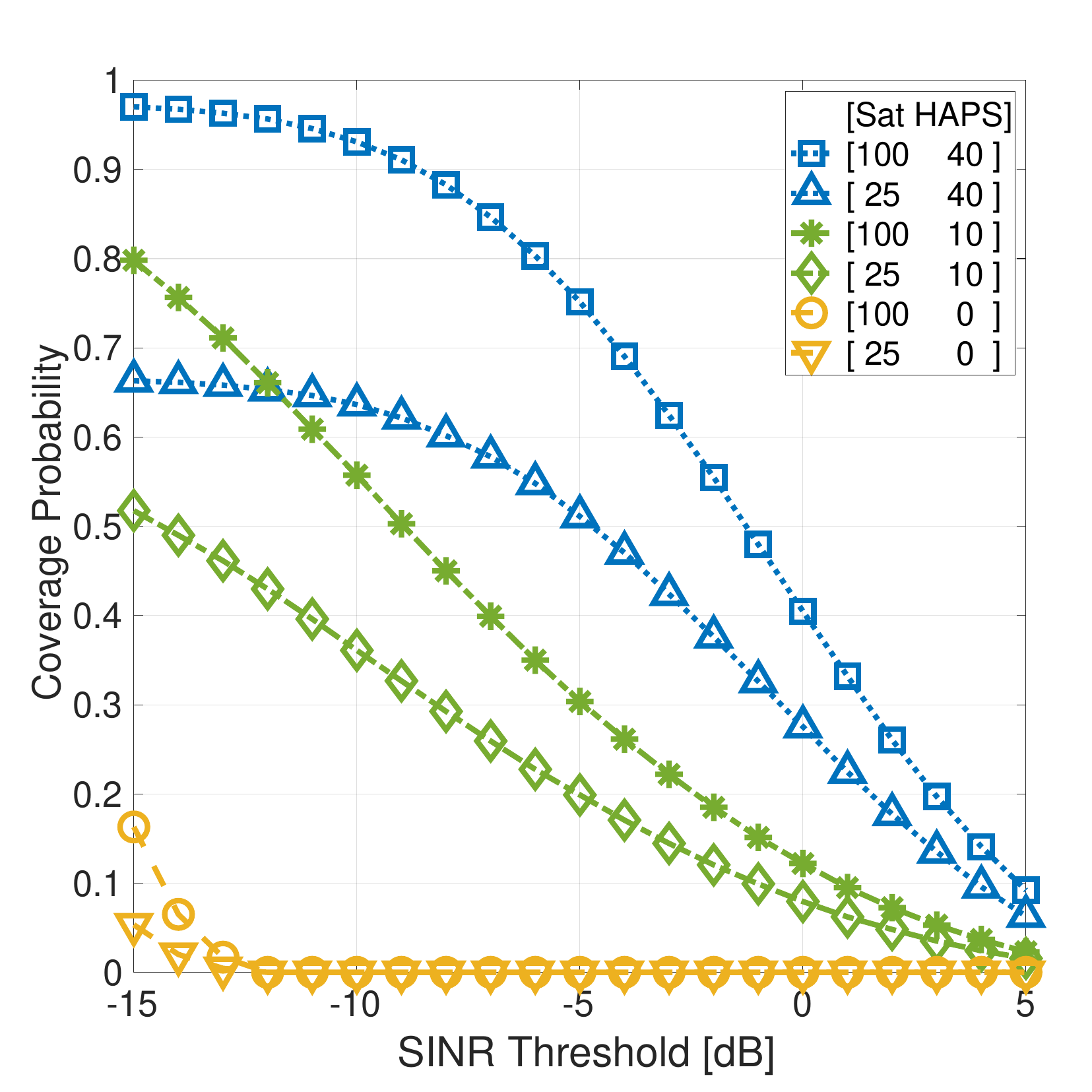}
	\caption{Coverage probability as a function of SINR threshold in HAPS-relayed cell-free networks.}
	\label{fig:coverage}
\end{figure}

\subsubsection{Insights}
{\color{black} As shown in Fig.~\ref{fig:coverage}, when the threshold exceeds $-10$~dB, satellites have to employ HAPS as a relay to communicate with the ground users. 
To enhance coverage probability, it is more effective to increase the deployment of HAPS while maintaining a constant number of satellites, as opposed to deploying more satellites while keeping the number of HAPS constant. This highlights the significant role that HAPS play in increasing coverage probability,  especially for smaller SINR threshold. The convergence trend of the coverage probability is mainly determined by the number of HAPS, whereas the number of satellites has a marginal impact on convergence. This indicates that for large SINR thresholds, HAPS failing to provide coverage for users is a major contributor to the communication outage.}

\subsection{Downlink and Uplink Capacity in HAPS-Based IAB Architectures}

\subsubsection{System Model}
\begin{figure}[ht]
	\centering
	\includegraphics[width=0.9\linewidth]{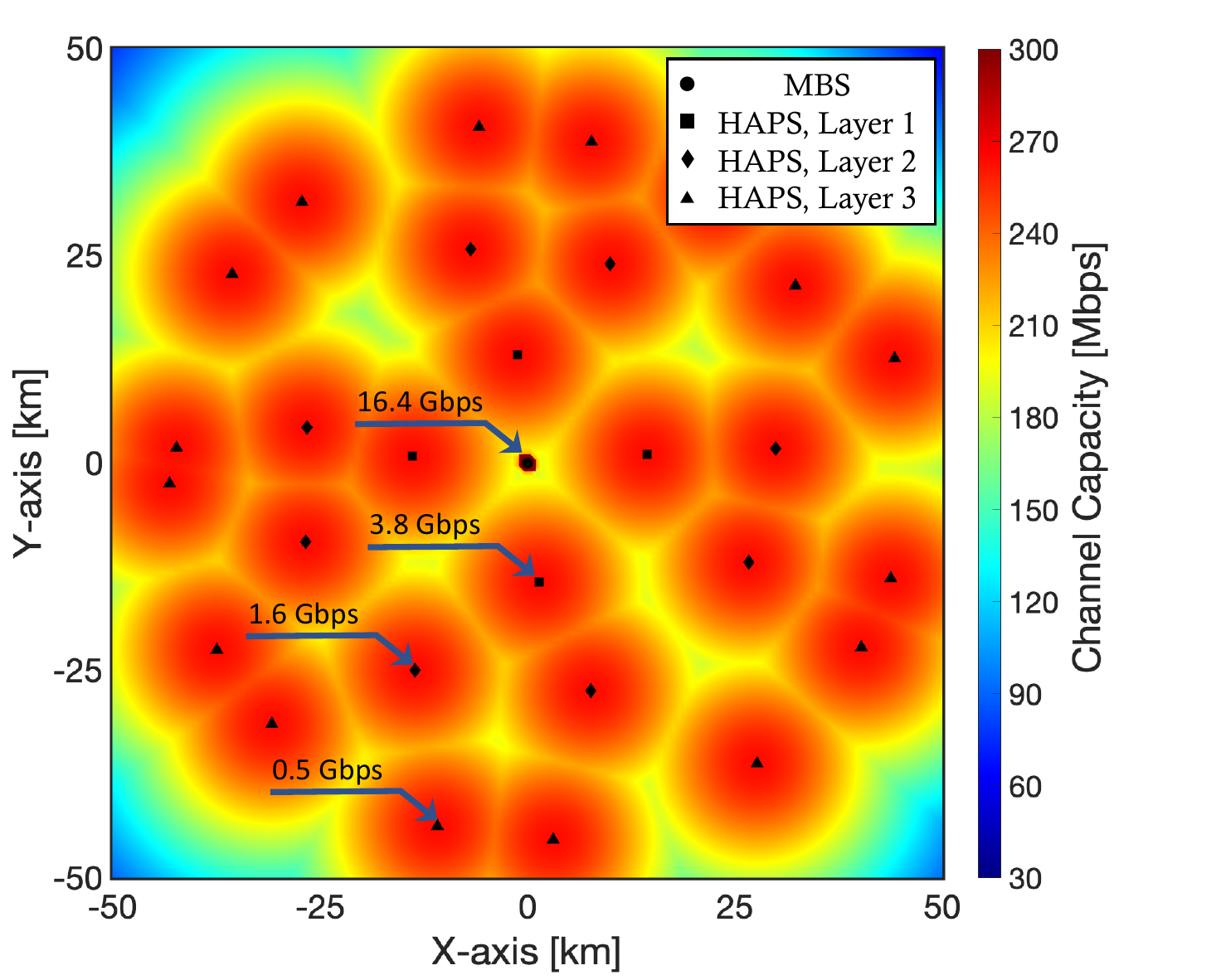}
	\caption{Heat map of downlink channel capacity in HAPS-based IAB architectures.}
	\label{fig:capacity}
\end{figure}

A three-layer IAB network consisting of $28$~HAPS and $1$~MBS is analyzed. In a region with a radius of $50$~km, the MBS is positioned in the center of the disk with a height of $10$~m and has the same antenna gain and transmit power as HAPS. 
As shown in Fig.\ref{fig:capacity}, the HAPS are divided into $3$ layers according to their distance from MBS in the IAB network. 
According to the distance to the MBS, the HAPS are categorized into three tiers, with $4$, $8$, and $16$ HAPS located in layers $1$, $2$, and $3$ of the IAB network, respectively.  
The distribution of users follows the same model as in subsection~\ref{simulation2}
{\color{black} All the links involved in this scenario utilize mmWave transmission with the parameters listed in Table~\ref{table}, and utilize orthogonal sub-bands and narrow-beam transmission to reduce interference. The channel model of the HAPS-HAPS/user/MBS link is analogous to the one used in \cite{wang2022stochastic}, whereas the MBS-user link is similar to the one used in  \cite{alzenad2019coverage}.}

\subsubsection{Metric and Insights}
{\color{black} {\color{black} Fig.~\ref{fig:capacity} illustrates the heat map representing the downlink channel capacity in relation to the HAPS-user link, as calculated according to Shannon's Law.} With a reasonable allocation of resource blocks, that is, channel capacity in this case, the entire IAB network can provide almost the same channel capacity for users in the region. {\color{black}The uplink aggregate data rate of MBS and HAPS located in different layers are also annotated in Fig.~\ref{fig:capacity}. HAPS closer to the MBA are allocated more resource blocks and thus have a larger uplink aggregate data rate. Noted that the uplink data volume is considerably smaller than that of the downlink due to the lower transmission power of the users.}
It is straightforward to notice that HAPS greatly extends the coverage region of the MBS. For instance, in areas with high demand for high-speed data, such as hotspots, the use of HAPS can ease network congestion and enhance overall network performance. }

\textcolor{black}{\subsection{Summary}
In what follows, we summarize the main takeaway insights
\begin{itemize}
    \item In UAV ad-hoc networks, introducing a HAPS as the center for routing scheduling and providing additional direct links can effectively reduce the latency for long-hop routing. Short-hop routing in densely deployed UAV ad-hoc networks does not rely much on UAV-HAPS direct links.
    \item In HAPS-centered networks, choosing a cell-free architecture over a traditional cellular structure sacrifices energy efficiency, but it also provides users with more association options. In cell-free architecture, increasing the number of HAPS will increase the energy efficiency gains.
    \item In satellite cell-free networks, the usage of HAPS as relays can significantly improve the coverage performance of satellite-ground communication.
    \item HAPS-aided IAB architecture can greatly expand the MBS service region. HAPS can provide almost equal downlink channel capacity for users in this region. The channel capacity of the backhaul uplink is divided into multiple levels according to the distance from HAPS to the MBS.
\end{itemize}}

\section{Conclusion}
We presented in this paper the main communication links between HAPS and other NTN platforms, their advantages, and their challenges. We also presented prospective network architectures in which HAPS plays an indispensable role in the future NTNs such as ad-hoc, cell-free, and IAB.
We provided comprehensive performance insights when using HAPS in the prospective architectures with the most suitable communication link.
The insights showed that in UAV ad-hoc networks, using HAPS as the center for routing scheduling and providing additional direct links effectively reduced the latency for long-hop routing.
In HAPS-centered networks, cell-free architecture decreased energy efficiency but provided users with more association options. In cell-free architecture, increasing the number of HAPS increased significantly the energy efficiency gains. In satellite cell-free networks, the usage of HAPS as relays significantly improved the coverage performance of satellite-ground communication.
Finally, HAPS-aided IAB architecture greatly expanded the MBS service region.

\bibliographystyle{IEEEtran}
\bibliography{references}

\section*{biographies}
\vspace{-0.5cm}
\begin{IEEEbiographynophoto}
{Zhengying Lou} is a Ph.D. student at King Abdullah University of Science and Technology (KAUST), Thuwal, Saudi Arabia. She received her B.S. degree in Communication Engineering from University of Electronic Science and Technology of China in 2021. In 2022, she received her M.Sc. degree in Electrical and Computer Engineering from KAUST. Her current research interests include stochastic geometry and wireless networks.
\end{IEEEbiographynophoto}
\vspace{-0.5cm}
\begin{IEEEbiographynophoto}
{Baha Eddine Youcef Belmekki} received the B.S. degree in electronics, and the M.Sc. degree in wireless communications and networking from the University of Science and Technology Houari Boumediene, Algiers, Algeria, in 2011 and 2013, respectively, and the Ph.D. degrees in wireless communications and signal processing from the National Polytechnic Institute of Toulouse (INPT), Toulouse, France, in 2020. He is currently a Postdoctoral Research Fellow with the Communication Theory Laboratory, KAUST. His research interests include non-orthogonal multiple access systems and stochastic geometry analysis of vehicular and aerial networks.
\end{IEEEbiographynophoto}
\vspace{-0.5cm}
\begin{IEEEbiographynophoto}
{Mohamed-Slim Alouini} [S’94, M’98, SM’03, F’09] received his Ph.D. degree in electrical engineering from the California Institute of Technology, Pasadena, in 1998. He served as a faculty member at the University of Minnesota, Minneapolis, then at Texas A$\&$M University at
Qatar, Doha, before joining KAUST as a professor of electrical engineering in 2009. His current research interests include the modeling, design, and performance analysis of wireless communication systems.
\end{IEEEbiographynophoto}

\end{document}